\documentclass[11pt,a4paper]{article}
\usepackage[utf8]{inputenc}
\usepackage[T1]{fontenc}
\usepackage{amsmath}
\usepackage{amsfonts}
\usepackage{amssymb}
\usepackage{graphicx}
\usepackage{subfig}
\usepackage[left=2cm,right=2cm,top=2cm,bottom=2cm]{geometry}
\usepackage{authblk}
\usepackage{xcolor}

\title{Study of Exospheric Neutral Composition of Mars observed from Indian Mars Orbiter Mission}

\author[$\star$ $\#$]{Kamsali Nagaraja}
\author[$\star$]{Praveen Kumar Basuvaraj}
\author[$\star$]{S. C. Chakravarty}
\author[$\dagger$]{Praveen Kumar K}

\affil[$\star$]{Department of Physics, Bangalore University, Bengaluru 560056 India}
\affil[$\dagger$]{Indian Space Research Organisation Headquarters, Bengaluru 560231 India}
\affil[$\#$]{Corresponding author: kamsalinagaraj@bub.ernet.in}
\date{}

\begin{document}

\maketitle

\begin{abstract}
The raw exospheric composition data of Mars for the period September~2014 to October~2015 has been retrieved and analysed using the observations carried out by Mars Exospheric Neutral Composition Analyser (MENCA) payload on-board the Mars Orbiter Mission (MOM) launched by India on 5~November~2013. The state parameters viz. latitude, longitude, altitude and solar zenith angle coverage of the partial pressures of different exospheric constituents are determined and assigned to enrich the usefulness of data with orbit-wise assimilation particularly between 250 and 500 km altitude range of main interest. Apart from getting the results on mean individual orbits’ partial pressure profiles during the northern winter, the variations of total as well as partial pressures are also studied with respect to the distribution of major atmospheric constituents and their dependence on solar activity. In particular, $CO_2$ and $O$ variations are considered together for any differential effects due to photo-dissociation and photo-ionisation. The results on gradual reduction in densities due to the decreasing daily mean sunspot numbers and strong response of $CO_2$ and $O$ pressures to solar energetic particle events like that of 24~December~2014 following a solar flare were discussed in this paper.
\end{abstract}


\section{Introduction}
Early measurements of atmospheric densities and chemical composition close to the surface of Mars were carried out using earth-based remote sensing techniques of high resolution optical spectrometers incorporating the Doppler shift method. These measurements in visible and IR bands indicated $CO_2$ as the dominant constituent with the total atmospheric pressure of about $\approx$7 mb which were later found to be corroborated with the Viking 1 and 2 lander spacecraft results (Moroz et al., 1998 and references there in). Similarly from earth-based spectral line measurements, the Martian surface-air concentrations of water vapour ($H_2O$), molecular oxygen ($O_2$), carbon monoxide ($CO$) and isotope ratios of oxygen ($O$) and carbon ($C$) were determined (Kaplan et al.,1969; Barker et al., 1972; Carleton and Traub, 1972; Young and Young, 1977). During the initial years of space exploration, optical spectrometers in UV and IR bands were used in e.g., Mariner 9 orbiter mission resulting in the detection of hydrogen corona and the confirmation of small densities of ozone ($O_3$) in the polar region of Mars (Lane et al., 1973; Barth and Dick, 1974). There were similar observations of lower atmospheric constituents from other spacecraft missions including Mars 3 \& 5 and Mariner 6 \& 7. While the efforts to measure the atmospheric parameters including the densities of the gaseous constituents near the surface of Mars in the pre-Viking period i.e., before 1976, have provided some basic information, there was a need to carry out further observations particularly related to the spatial distribution of the atmospheric constituents in the thermosphere and exosphere of Mars. Such an opportunity was provided by the launch of two Viking spacecraft missions during September~1976 to carry out the observations using mass spectrometers both on the entry probe as well as on the lander. The first in-situ data of the Martian upper atmospheric composition, densities and temperatures were obtained from this twin Viking mission (Hanson et al., 1977; Nier and McElroy, 1977; Owen et al., 1977; Hanson and Mantas, 1988). This was followed by Phobos 2 (1989), Mars Global Surveyer (MGS, 1997) and the Mars Express (MEX, 2004) missions with main emphasis to study the interaction of solar wind charged particles with Mars, global mapping of the magnetic field and estimating escape rates of ions from Mars (Sagdeev and Zakharov, 1989; Acuna et al., 1999; Barabash et al., 2007). The Mars Pathfinder mission (1997) was the first successful lander-rover mission with similar findings as that of Viking (Magalhaes et al., 1999). Mars Odyssey (MO, 2002) and Mars Exploration Rovers Spirit and Opportunity (MER, 2004) searched for the surface features affected by flow of water and the atmosphere close to the ground (Mangold et al., 2004; Squyres et al., 2006). More recent missions like Mars Reconnaissance Orbiter (MRO, 2006), Phoenix lander (2008) and Mars Science Laboratory (MSL, 2012) revealed hemispherical differences in the variation of precipitable water (Smith et al., 2009), role of recent degassing due to volcanic activity in the evolution of modern atmosphere of Mars (Niles et al., 2010) and enhanced $D/H$ (Deuterium/Hydrogen) ratio in the clay minerals pointing to a longer history of hydrogen escape and hence water (Mahaffy et al., 2015). The quadrupole mass spectrometer (QMS) as part of the instrument suite on the MSL’s Curiosity rover measured the chemical concentrations and isotopic fractions of volatile compounds. The surface atmospheric composition of Mars consists of about 96\% carbon dioxide ($CO_2$) and small concentrations of nitrogen ($N_2$), argon ($Ar$), and trace amounts of oxygen ($O_2$), Water vapour ($H_2O$) and ozone ($O_3$) (Mahaffy et al., 2013). The complete set of near surface meteorological data obtained from Viking landers (1976) to the Curiosity rover (2012) have been analysed and results have been consolidated in terms of diurnal, seasonal and inter-annual variations of meteorological parameters including dust storms over a span of more than 20 Martian years (Martinez et al., 2017). A similar analysis to characterise the thermosphere-exosphere system has not been possible due to paucity of observational data building up time sequences of the altitude and latitudinal profiles of the meteorological parameters and neutral/ionised gas concentrations (Bougher et al., 2014). 

Recently the Indian Mars Orbiter Mission (MOM, 2014) and the Mars Atmosphere and Volatile Evolution (MAVEN, 2014) from NASA have been launched for the study pertaining to the evolution and escape of atmospheric/ionospheric constituents. MAVEN makes additional observations of the constituents in the fringe of the mesosphere. It is an important point to note that till September~2014 when MOM and MAVEN arrived at Mars, the data from the Viking mission remained the only in-situ measurements of neutral and ion composition of the thermosphere and its fringe region just touching the exosphere of Mars. The main purpose of this paper is to study the spatial and temporal distribution of atmospheric composition of the thermosphere-exosphere region of Mars which is essential to understand the diurnal, latitudinal, seasonal and solar activity driven variations and the escape of neutral/ion atmospheric species from Mars. The in-situ observations carried out by MENCA (Mars Exospheric Neutral Composition Analyser) and NGIMS (Neutral Gas and Ion Mass Spectrometer) on board MOM and MAVEN orbiter missions have opened up new vistas to further characterise the thermosphere-exosphere of Mars. The primary objective is therefore to reduce the gap in knowledge about the variations of atmospheric constituents in the exobase to exosphere of Mars using recent data from MENCA. 

\begin{figure}[h]
	\vspace*{0cm}
	\centering
	\makebox[0pt]{%
	\includegraphics[width=0.8\paperwidth]{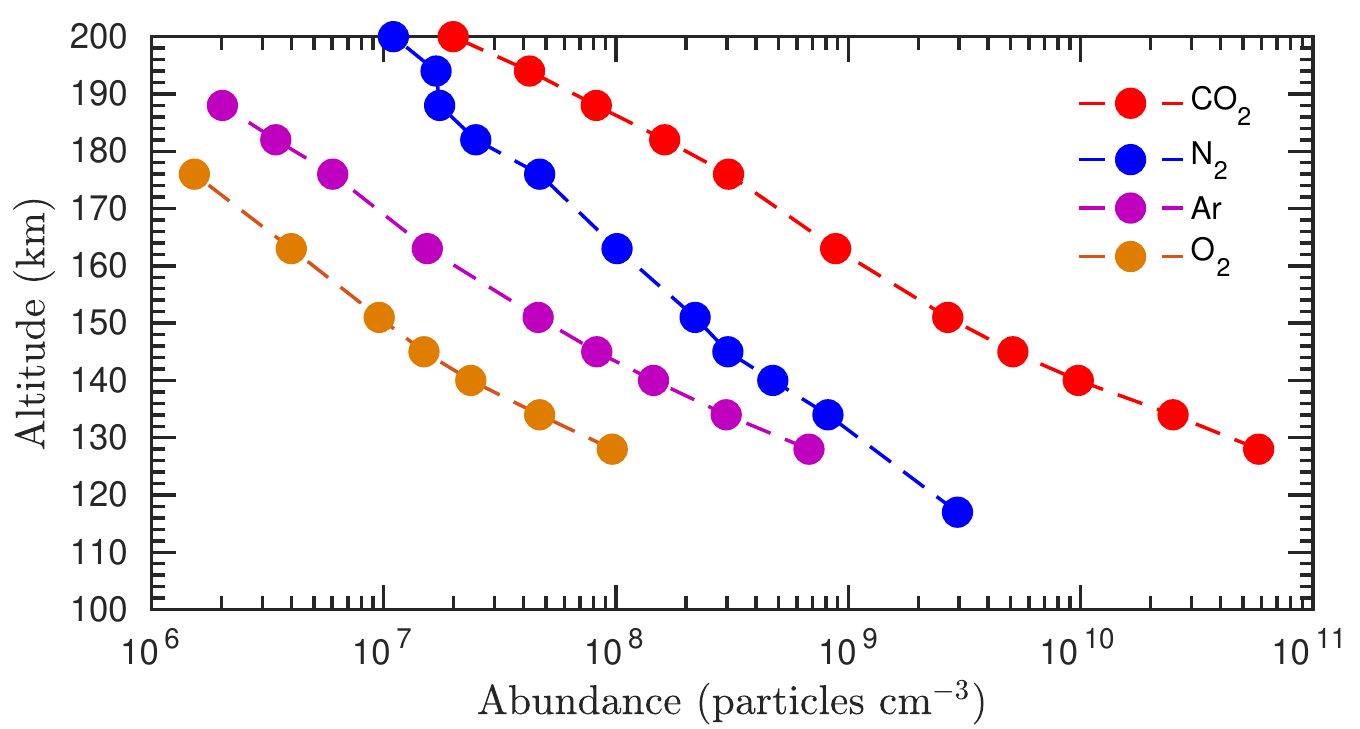}}
	\caption{Neutral composition number density profiles in the thermosphere and exobase regions of Mars as measured by Viking 1 lander (Nier and McElroy, 1977). The lander covered SZA of 41$^\circ$-44$^\circ$ and ground co-ordinates in latitude and longitude of 14$^\circ$-17$^\circ$ and 302$^\circ$-306$^\circ$  respectively.\label{overflow}}
\end{figure}

\section{Early measurements of the upper atmosphere of Mars}
The first in situ observations providing the vertical profiles of neutral gases, ion, electron densities and temperatures of the atmosphere of Mars were carried out by the Viking 1 and 2 landers when these descended down to the surface of Mars. 

The neutral mass spectrometric data on the composition of the upper atmosphere of Mars indicates that the main constituents of the Martian atmosphere are carbon dioxide ($CO_2$), Argon ($Ar$), molecular nitrogen ($N_2$), carbon monoxide ($CO$), and with photo-chemically produced atomic oxygen ($O$) (Nier and McElroy, 1977). The surface level volume mixing ratios of $CO_2$, $Ar$ and $N_2$ are 0.960, 0.0193 and 0.0189 respectively (Mahaffy et al., 2013). The altitude variations of some important atmospheric constituents above the well mixed atmosphere, i.e., above the mesosphere and into the thermosphere, where the diffusion of individual atmospheric species dominate in comparison to turbulent mixing (Haberle et al., 2002), are shown in Figure~1 as an example taken from Viking results (Nier and McElroy, 1977).

During the period of atmospheric observation while in descent from 200 km, the lander covered SZA of 41$^\circ$-44$^\circ$ and ground coordinates in latitude and longitude of 14$^\circ$-17$^\circ$ and 302$^\circ$-306$^\circ$  respectively (Withers and Lorenz, 2002). So the data on altitude variation of the gaseous constituents shown in Figure 1 are taken under similar solar radiation condition and within a marginal variation of 4 degrees of latitude and longitude. These results were later extended from 200 to 300 km through scale height extrapolations (Hanson et al., 1977). After the operation of Viking, a number of orbiter, lander and rover missions have enriched the information about Mars’s atmospheric phenomena mainly in its meteorological context. Based on theoretical considerations, Mars atmospheric general circulation models such as Mars Climate Database (MCD) and Mars Thermospheric General Circulation Model (MT-GCM) have been developed extending into the exosphere but validated mainly using the in-situ observations from Viking (Lewis et al., 1999; Bougher et al., 2012). Another advantage of MOM/MAVEN data is that it helps determining the solar activity effects as the observations have been taken during the period of moderately high solar activity condition as compared to Viking probe measurements. 

The present work therefore involves conversion of the available near-raw data of MENCA experiment into a calibrated data set with associated tags of time, altitude, latitude, longitude and solar zenith angles for detailed analysis to derive the exospheric composition and its variability. 

\section{Observation and Data Analysis}
The Indian Mars Orbiter Mission (MOM) was launched on 5~November~2013. The probe arrived in Mars on 24~September~2014 into an highly eccentric orbit of 422~km x 76,993~km with an orbital period of $\approx$~67~hours. Later manoeuvres during December~2014 brought down the periareion altitude to around 262~km (Bhardwaj et al., 2016). One of the payloads of MOM is MENCA for the measurement of total atmospheric pressure and partial pressures of various atmospheric constituents covering 1-300 amu (Bhardwaj et al., 2015). The data sets used for this paper are obtained when MENCA was operated in the mass range of 1-100 amu. Raw data of MENCA payload were made to public for the period 24~September~2014 to 23~September~2017 through the ISRO Space Science Data Center (ISSDC). This archived data consists of total and partial pressure values in the units of Torr with time resolution of 12 to 30 seconds near periareion. Before this data can be used for scientific studies, it needs further processing in terms of calibration factors and ancillary data like latitude, longitude, altitude and solar zenith angle.

\begin{table}[h]
	\vspace*{0cm}
	\caption{Data availability sample; giving MOM orbit number and number of data files along with their time coverage}
\begin{center} 
 \begin{tabular}{|c|c|c|c|}
		\hline
		Orbit \# & Observation start time & Observation end time & Total files \\			
		\hline
		0031    & 2014-10-17 11:01:17	& 2014-10-19 18:43:23	& 28 \\
		0032	& 2014-10-20 02:01:18	& 2014-10-21 06:38:39	& 10 \\
		0034	& 2014-10-28 05:41:41	& 2014-10-30 20:32:55	& 16 \\
		0035	& 2014-11-01 23:16:21	& 2014-11-02 14:01:41	& 04 \\
		0036	& 2014-11-02 16:26:04	& 2014-11-05 07:11:51	& 16 \\
		0038	& 2014-11-05 10:13:05	& 2014-11-08 00:40:58	& 10 \\
		0039	& 2014-11-08 03:34:50	& 2014-11-10 18:08:43	& 12 \\
		0040	& 2014-11-11 19:03:42	& 2014-11-13 11:57:10	& 10 \\
		\hline	
\end{tabular}
\end{center}
\end{table}

The data is identified with respect to different orbit numbers of MOM for aforementioned period covering 88 orbits. The raw data of each orbit is included in a number of file pairs, each pair containing: (a) total atmospheric pressure (in Torr) and (b) partial pressures (in Torr) of the constituents for different time intervals. The individual files for partial pressures contain the data in a continuous time sequenced array form, which is converted into tabulated columnar format by marking each column for: (i) time of observation in UTC, (ii) corresponding 9-values of partial pressures for each constituent with amu between 1-100 and (iii) total pressures time synchronised with partial pressure measurement times using the second file in the pair. Table 1 shows an example of information on number of files containing data in a streaming time sequence mode for different orbits. ISSDC has also provided the specific SPICE (Spacecraft Planet Instrument C-matrix Events) kernel files for one year period of observations consistent with the internationally standardized service modules (Acton C.H. et al., 1996; 2017) required to compute the altitude, latitude, longitude and solar zenith angle values tagged to each time of observation so as to characterize and tabulate the pressure data with respect to these localization parameters. 

Further data processing involves: (a) combining each pair of file to convert into one file having the partial pressures (maximum pressure value out of 9-values of observation for each of the 100 amu bins) as well as the time synchronised total pressure. Thus 100 maximum pressure values are selected for each amu from 1-100 for the same time epoch constituting a single record for further analysis, (b) computing the altitude, latitude, longitude and solar zenith angle for each time epoch of observation using the kernels of SPICE system mentioned earlier and (c) conversion of raw partial pressure data using calibration information provided along with the science data for different amu values. The final calibrated partial pressure values along with the associated ephemeral and spatial information for each orbit is used for further analysis and scientific studies of the atmospheric constituents of Mars. After the above treatment of data, the MOM orbit-wise analysis is carried out to select the useful partial pressure values from the lowest altitude to about 500 km. This height limit is decided based on the requirement of our present study. The sum of the partial pressures of major gas constituents is also used to compute the total pressures for studying relative variation with time, altitude, etc. This method is found useful when the concentrations of certain constituents like water vapour, molecular and atomic hydrogen etc. show abnormally high values due to out-gassing. Such contamination due to out-gassing has been reported by the payload team (Bhardwaj et al., 2017). Statistical analysis is carried out for deriving the monthly mean values and standard errors at 95\% confidence interval.

\section{Results}
The MENCA in-orbit payload operation plan was governed by the science objective to study the atmospheric composition in the exosphere which has been constrained by the variation of periareion (260-430 km) for the first year (September~2014 to October~2015) of observation. However this periareion coverage of lowest altitudes is best suited to meet the scientific objectives compared to the situation of subsequent observations.

\begin{figure}[h]
	\vspace*{0cm}
	\centering
	\makebox[0pt]{%
	\includegraphics[height=8.5cm,width=0.8\paperwidth]{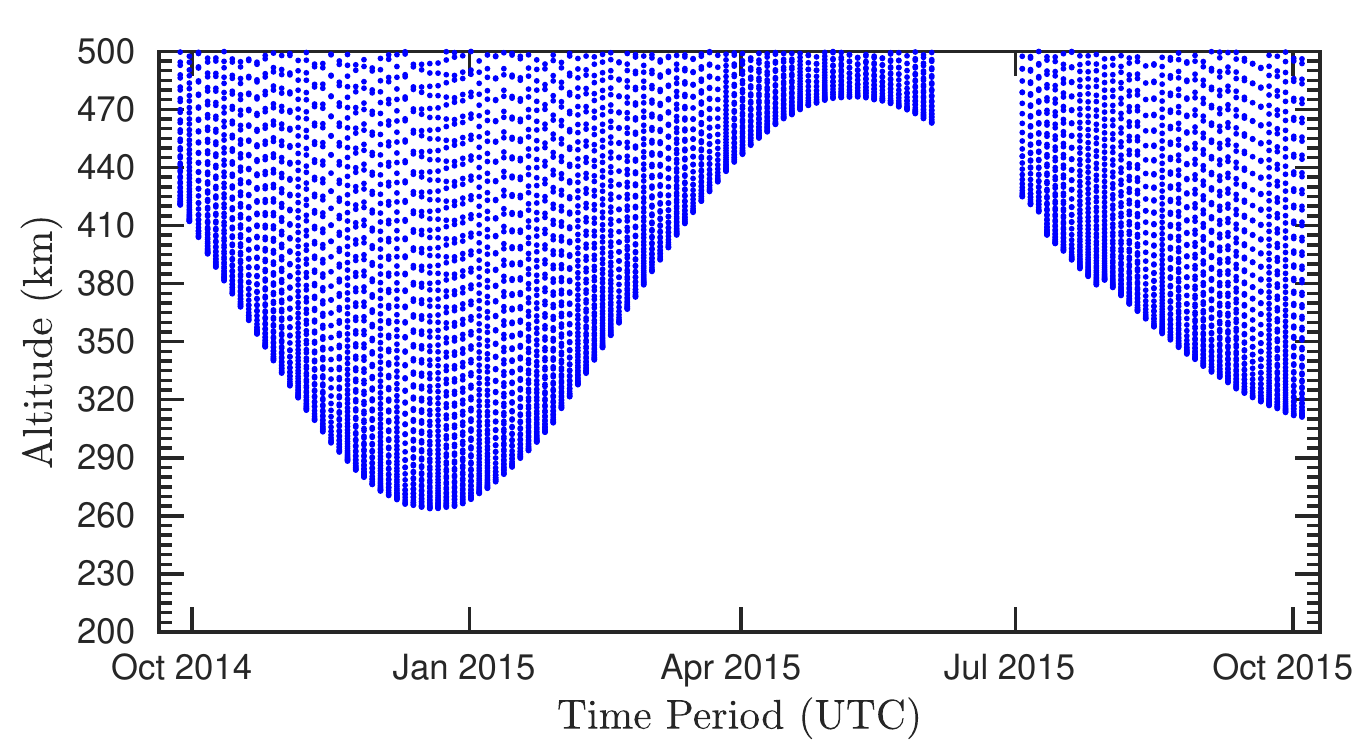}}
	\caption{Periareion altitude coverage of MENCA observations during October~2014 to October~2015\label{overflow}}
\end{figure}

\begin{figure}[h]
\vspace*{0cm}
	\centering
	\makebox[0pt]{%
	\includegraphics[height=8.5cm,width=0.8\paperwidth]{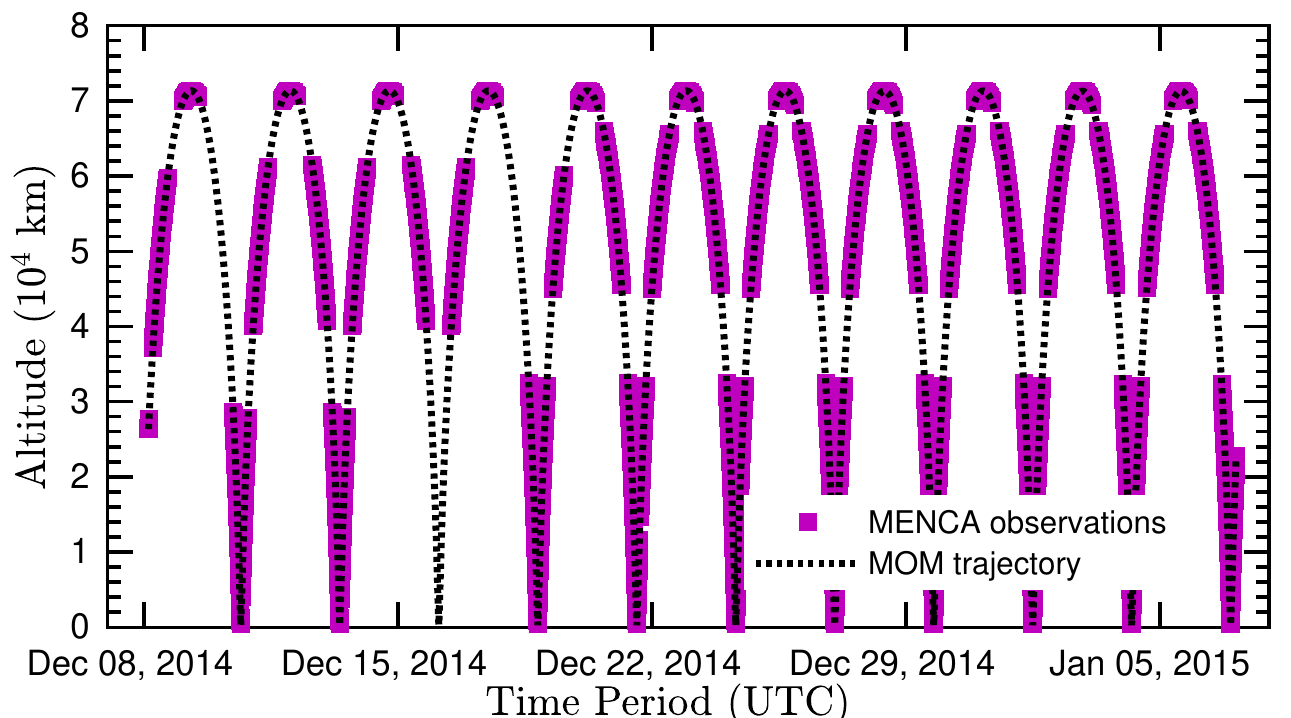}}
	\caption{MOM trajectory (dotted) and MENCA observations  (coloured) in each orbit during Dec~2014 \label{overflow}}
\end{figure}

Figure 2 shows this variation of the altitude of the periareion of MOM spacecraft during the first year of data collection. From 27~May~2015 to 01~July~2015, MOM was behind the Sun as viewed from the Earth and hence no observations were made during the communication black-out period. During each MOM orbit, the MENCA payload has been operated to cover the descending and ascending tracks near the periareion as well as at locations away from the periareion, mainly to serve as background information. 

Figure 3 shows the MOM tracks for the 11-orbits during December~2014 with the MENCA observation periods indicated by purple colour along these tracks. It can be seen that the observations have been taken at different sections of the track ensuring the altitudes near the periareion of MOM for observing the exosphere of Mars starting from about 260 km altitude. This height coverage of composition data is unique as there have been hardly any such in-situ measurements of Martian upper atmosphere deep into the exosphere.

The in-orbit mass spectrometric observations of MENCA payload are linked to the planetary co-ordinates in terms of altitude, latitude, longitude and SZA. These are determined using the auxiliary data files provided along with the pressure data of the atmospheric constituents. 

\begin{figure}[h]
	\vspace*{0cm}
	\centering
	\makebox[0pt]{%
	\includegraphics[width=0.8\paperwidth]{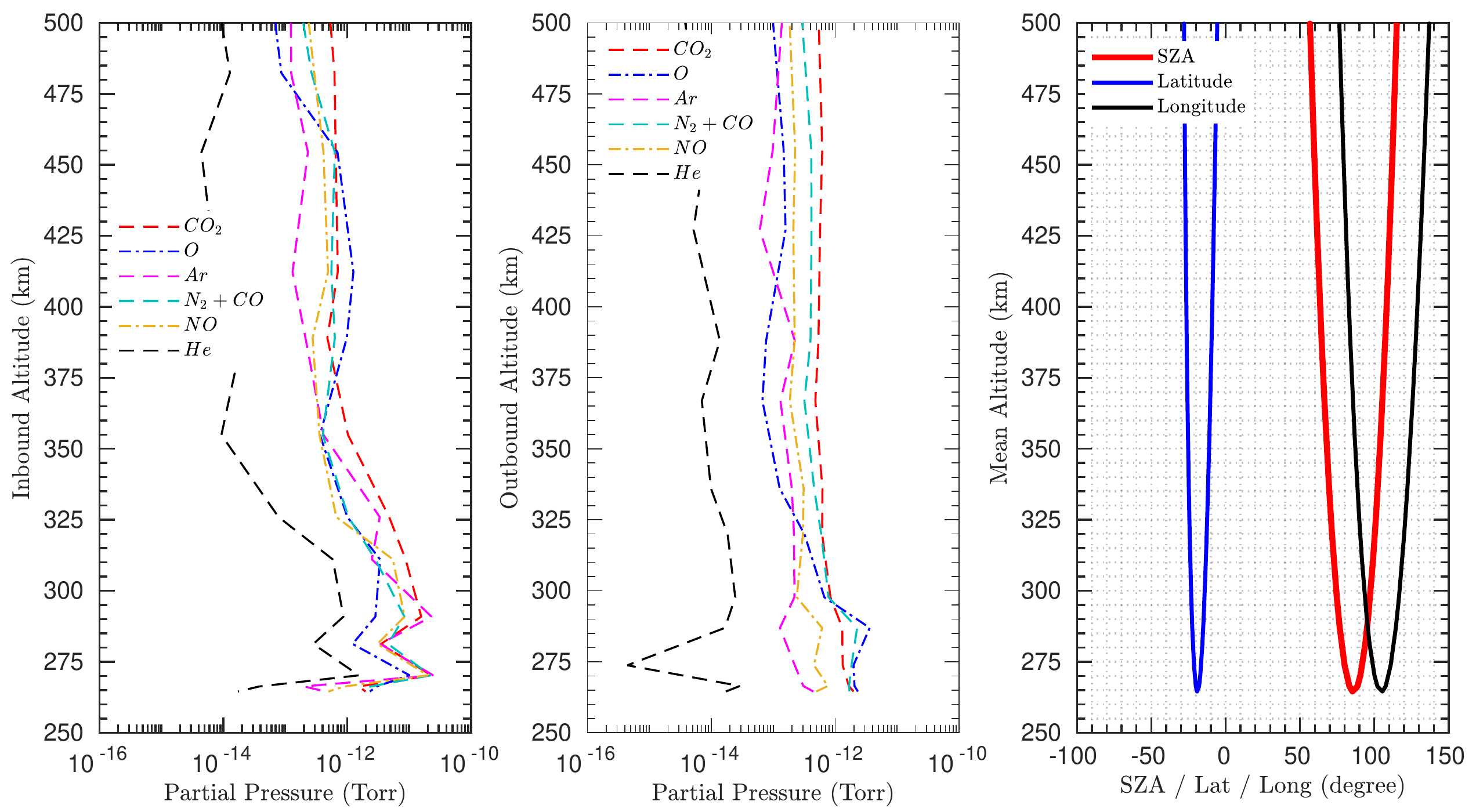}}
	\caption{Partial pressures of exospheric constituents from MENCA observations related to MOM orbit~$\#$55 on 24~December~2014, shown for both inbound and outbound separately along with the coverage of solar zenith angle, latitude and longitude.\label{overflow}}
\end{figure}

The MENCA dataset obtained following the above procedure has been analysed mainly to study the variation of partial pressures of gaseous composition. The exospheric profiles of these parameters are obtained for individual orbits specifically covering the region around periareion of MOM. A typical example of the altitude profiles of exospheric constituents derived from MENCA observations of the orbit $\#$54 for descending and ascending tracks of MOM is shown in Figure 4. Here we notice that pressure values drop with altitude by about 2-orders of magnitude between 260 and 500~km for descending track. However for the ascending part of orbit (a) the gradient is slower particularly between 260-350~km and (b) the absolute magnitude of this reduction in pressure is lower between 260 and 350 km during sunset as evident from the variation of SZA shown in the same figure. Combined together this decrease of pressure is slower with altitude compared to the Viking results below ~200 km due to the difference that above the thermosphere the gaseous elements follow diffusion equilibrium paths which are different for different species depending on their mass and temperature. It can also be seen that in the region of exosphere above 300~km, $CO_2$, $N_2$, $O$ and $O_2$ are the dominant atmospheric constituents. The figure also shows considerable reduction in atomic oxygen ($O$) density after sunset. This is because $O$ is mainly a product of photo-dissociation of $O_2$ and $CO_2$. As it is not possible to separate $N_2$ and $CO$ having the same amu of 28, the effective value of $N_2$ would be less due to the contribution of $CO$ which is a photo-dissociation product of $CO_2$ (Bhardwaj et al., 2017).

\newpage

\begin{figure}[h]
	\vspace*{0cm}
	\centering
	\makebox[0pt]{%
	\includegraphics[width=0.8\paperwidth]{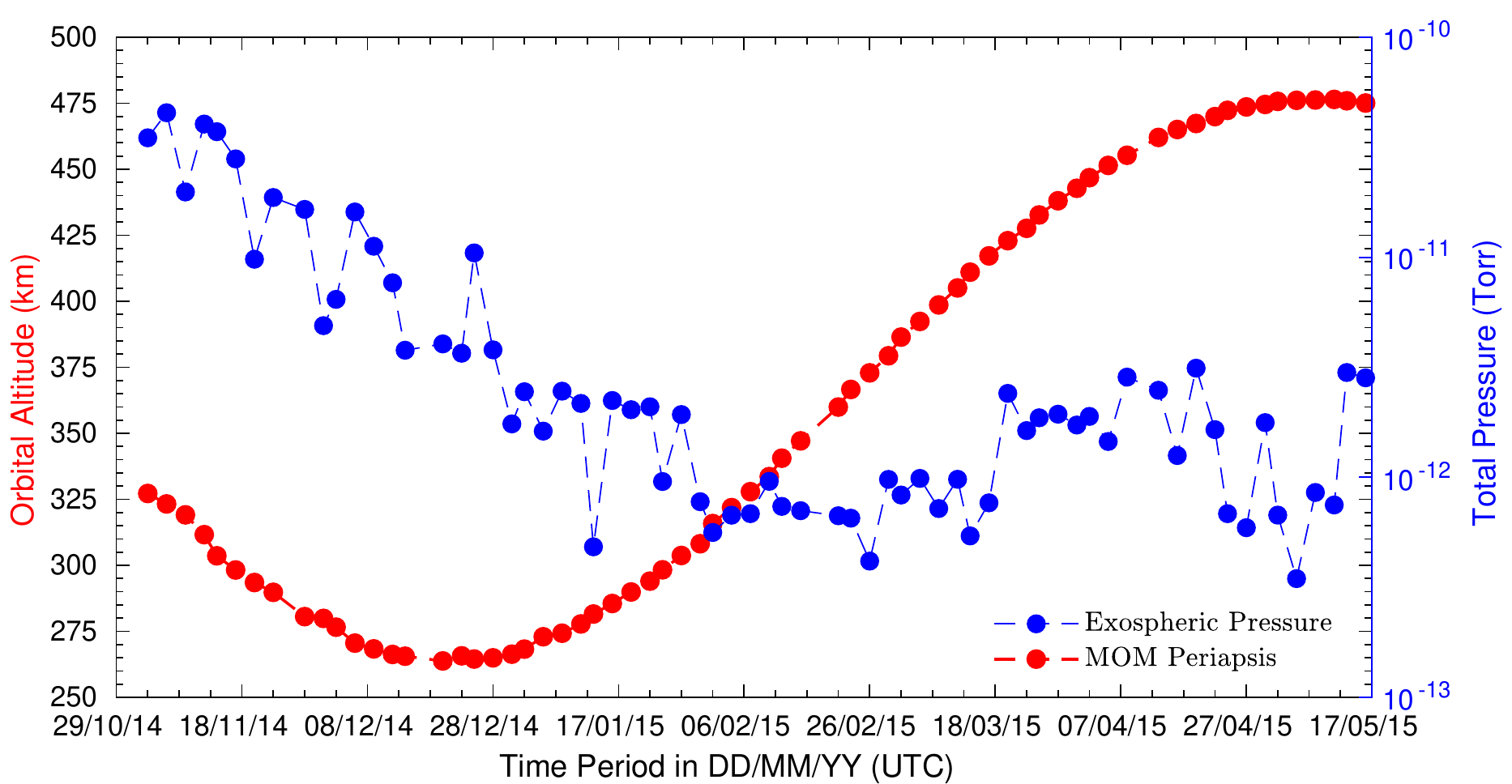}}
	\caption{Total pressure values estimated by summing partial pressures of a number of constituents for the minimum altitude of each MOM orbit ($\#$28-107) during the period October~2014 to May~2015.\label{overflow}}
\end{figure}

\begin{figure}[h]
	\vspace*{0cm}
	\centering
	\makebox[0pt]{%
	\includegraphics[width=0.8\paperwidth]{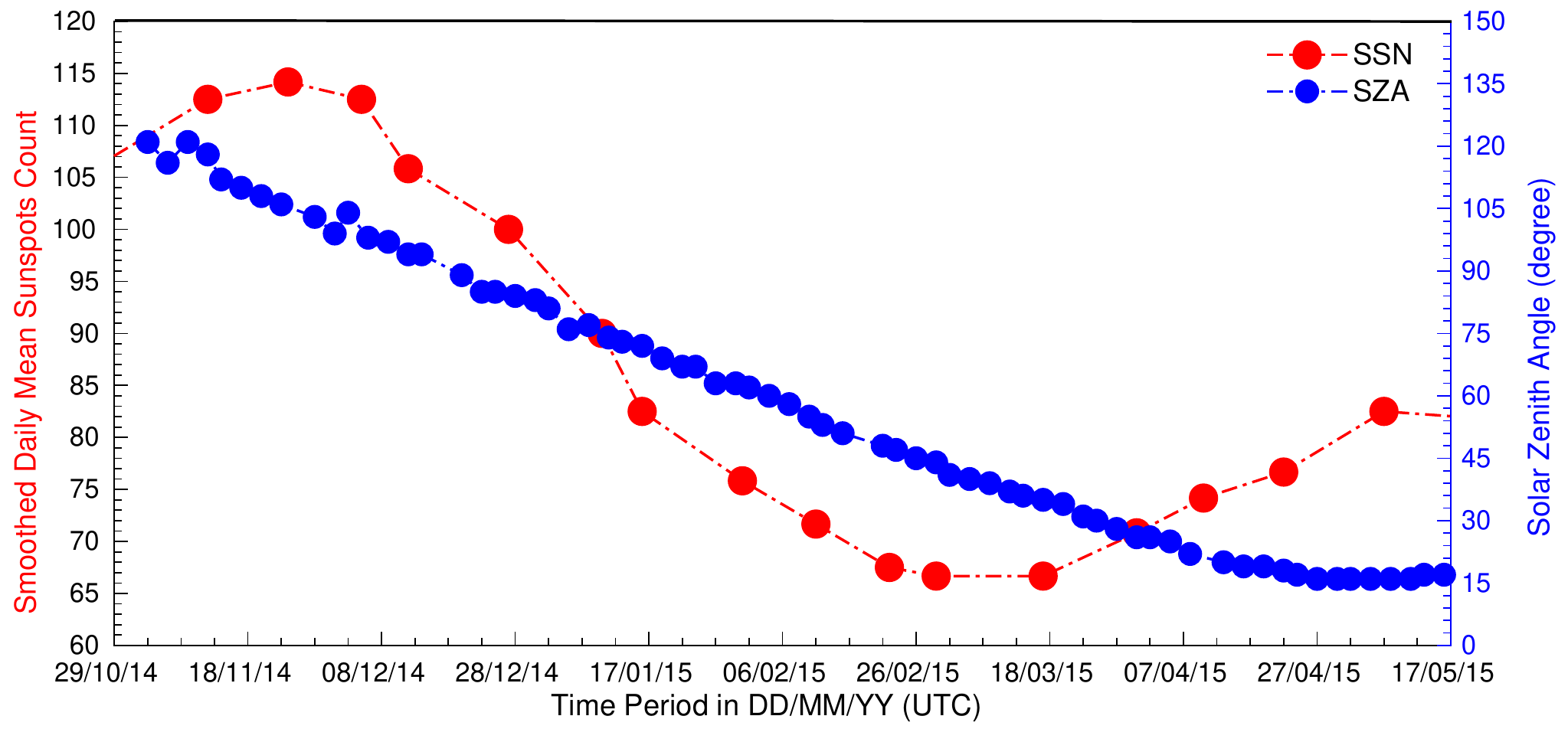}}
	\caption{Variation of smoothed daily mean sunspot numbers (SSN) and solar zenith angle for the same period as in Figure 5\label{overflow}}
\end{figure}

\begin{figure}[h]
	\vspace*{0cm}
	\centering
	\makebox[0pt]{%
	\includegraphics[height=13cm, width=0.8\paperwidth]{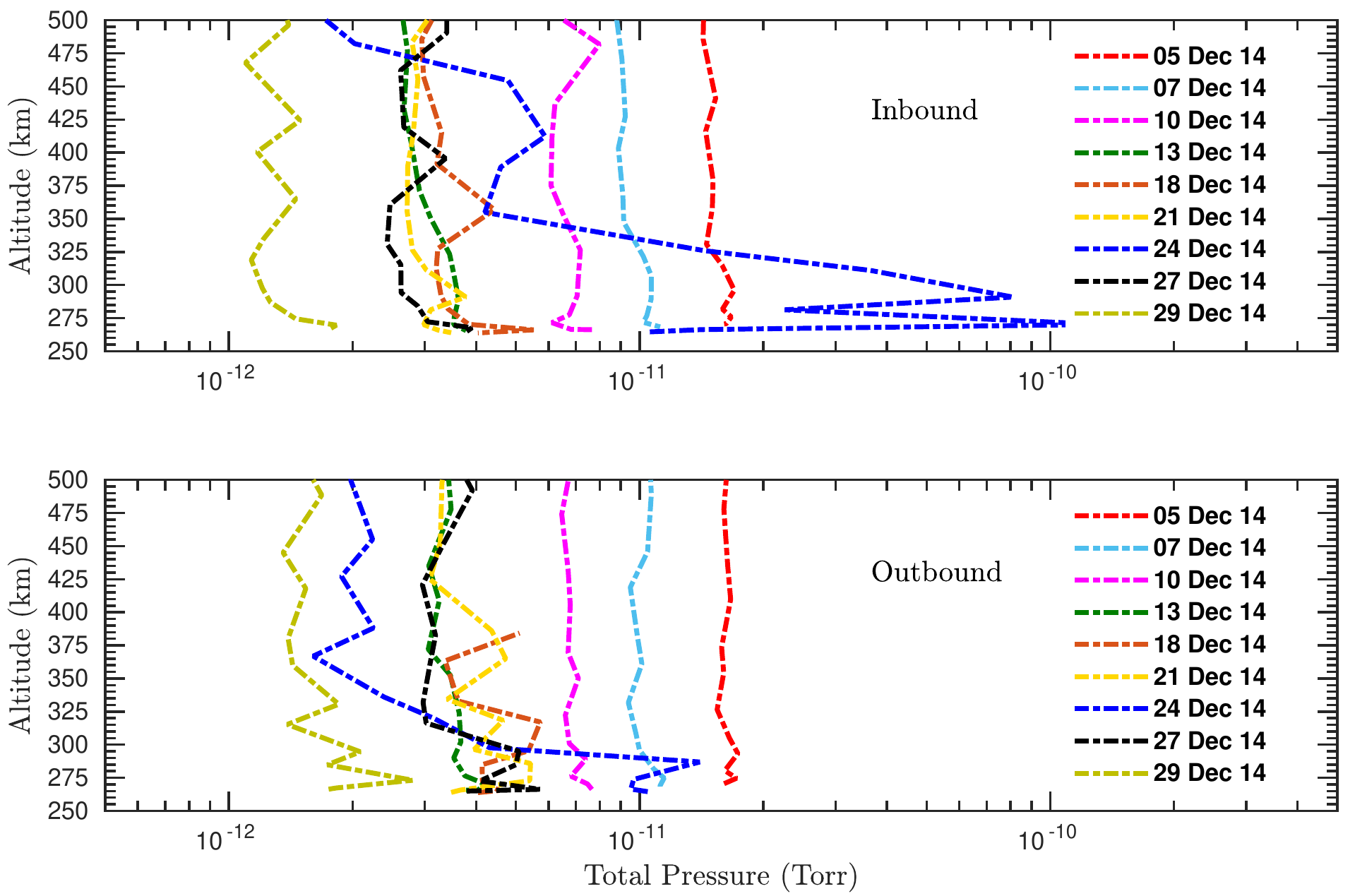}}
	\caption{Vertical profiles of total pressure for the inbound and outbound of MOM orbits during December~2014\label{overflow}}
\end{figure}

Hence, for the rest of the paper we would consider $CO_2$, $O$, $N_2$ and $Ar$ as the main elements to study the variation of exospheric composition. Also the total pressure is computed by summing up the partial pressures of all the major constituents as $CO_2$, $O$, $N_2$, $CO$, $NO$, $O_2$, $Ar$, $N$ and $He$. The contribution due to water vapour ($H_2O$), $H_2$ and $H$ are excluded from the sum as these constituents may be modulated by the degassing of the spacecraft.

The total pressure profiles for exospheric altitudes have been estimated from the available data per orbit and the results for both descending and ascending tracks are shown for the month of December, 2014 in Figure 7. It can be seen that, except for the anomalous pressure values on 24 December 2014 (reasons explained later in this paper), there is a progressive decrease in total pressure at all altitudes from beginning to end of the month and this decrease correlates well with the decrease in the smoothed daily mean SSN mentioned earlier. The large reduction of the total pressure within a month is difficult to explain without invoking the solar activity effects. By checking the variations of solar zenith angles, latitude and longitude, it is found that the effect of these is at best very marginal and not effective in bringing a change of this magnitude within a period of one month for these altitudes. 

\begin{figure}[h]
	\vspace*{0cm}
	\centering
	\makebox[0pt]{%
	\includegraphics[height=11cm,width=0.8\paperwidth]{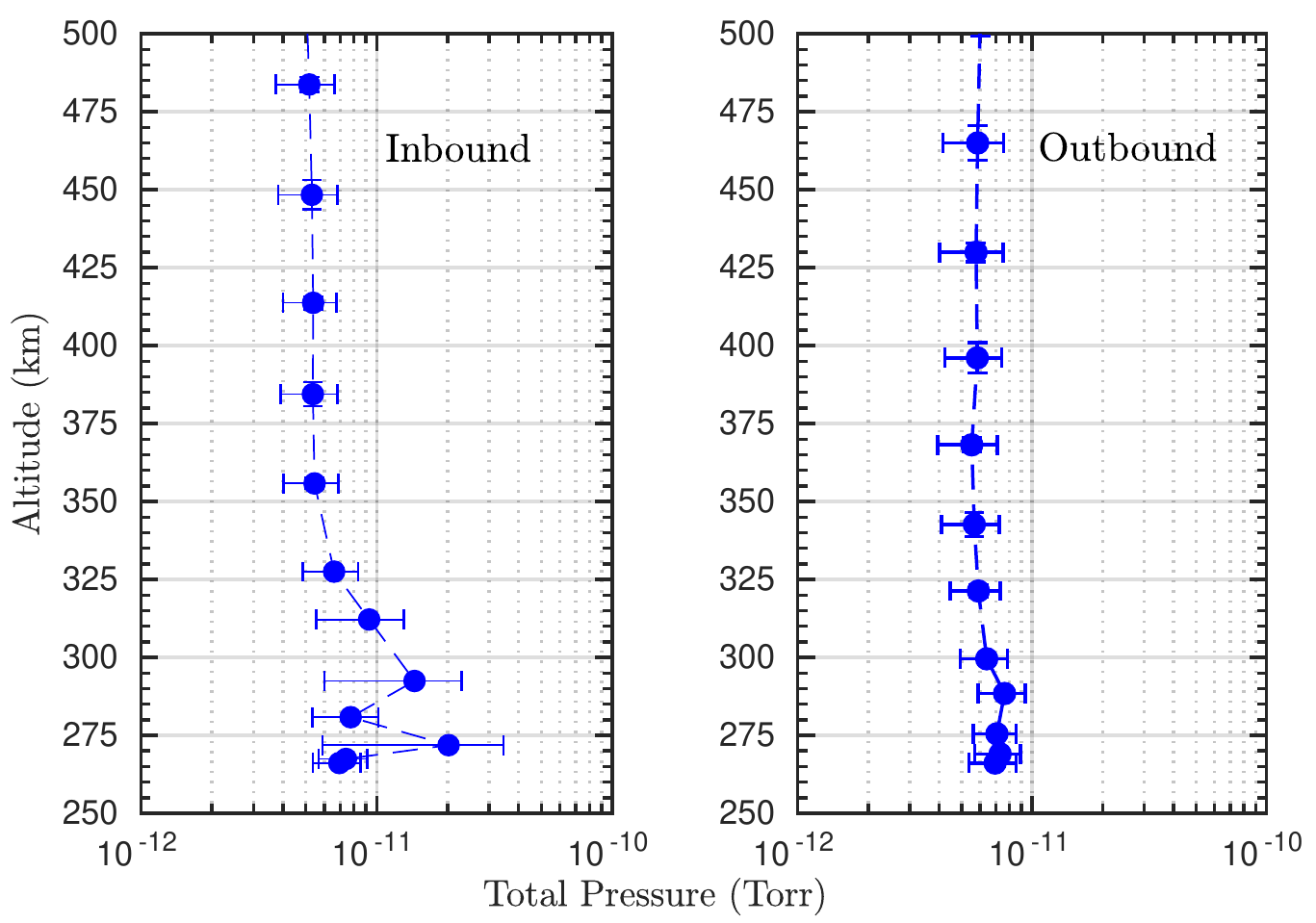}}
	\caption{Mean total pressure profiles with standard errors derived from all the measurements during December 2014 both for inbound and outbound.\label{overflow}}
\end{figure}

The changes in solar UV/X-ray fluxes arriving at Mars due to variations of solar activity have a strong modulating effect on exospheric pressures. These variations due to solar activity are found to be larger than the standard errors at 95\% confidence interval for both descending and ascending orbital paths as shown in Figure 8. Where the standard error of the means at 95\% confidence interval is given by m$\pm$(1.96 $\sigma/\sqrt{n}$) (m is the sample mean, $\sigma$ the standard deviation of mean and $n$ is the number of samples). These error bars are shown in the figure.

Only at two points between 250-300 km altitude for the descending track of the orbit the error bars are relatively large due to the anomalous pressures profile of 24 December, 2014. This point is addressed again later in this paper. Similar analysis is carried out separately for the partial pressures of $CO_2$ and $O$ by taking the mean pressure and altitude values of descending and ascending part of the orbital tracks. Figure 9 shows the individual days profiles of $CO_2$ and $O$ pressures for different days (pertaining to different orbit numbers) of December, 2014. From this figure the following points can be noted: (a) there is a considerable day to day variability of the altitude profiles of both $CO_2$ and $O$ with a consistent decrease of pressure at all atitudes with respect to advancing days in the period, (b) the $O$ pressure is generally higher than that of $CO_2$ before the anomalous event of 24 December, (c) On 24 December there is a large enhancement of $CO_2$ and $O$ pressures between 260-350 km when the solar proton fluxes are enhanced due to a moderate Coronal Mass Ejection (CME) event in progress during 20-25 December, 2014, (d) the anomalous pressure values of $CO_2$ are higher than that of $O$ at all altitudes and the $O$ pressures are lower compared to $CO_2$ pressures while recovering from the event after 24 December 2014. 

In order to explore the anomalous pressure enhancements of $CO_2$ and $O$, contours of these pressures with respect to time and altitude are plotted for periods between 21-27 December and 18-27 December respectively. These figures depict the following (a) $CO_2$ pressures started increasing anomalously from 22 December and reached 2 peaks by 24 December one around 275 km and the other around 295 km, (b) atomic oxygen ($O$) started increasing from 19 December and reached one strong peak around 275 km between 22-23 December, (c) the height range of the effect of pressure increase for $CO_2$ continued above 360 km but for $O$ it tapered around 330 km. 

\begin{figure}[h]
	\centering
	\begin{minipage}[b]{0.4\paperwidth}
		\includegraphics[height=6cm,width=\textwidth]{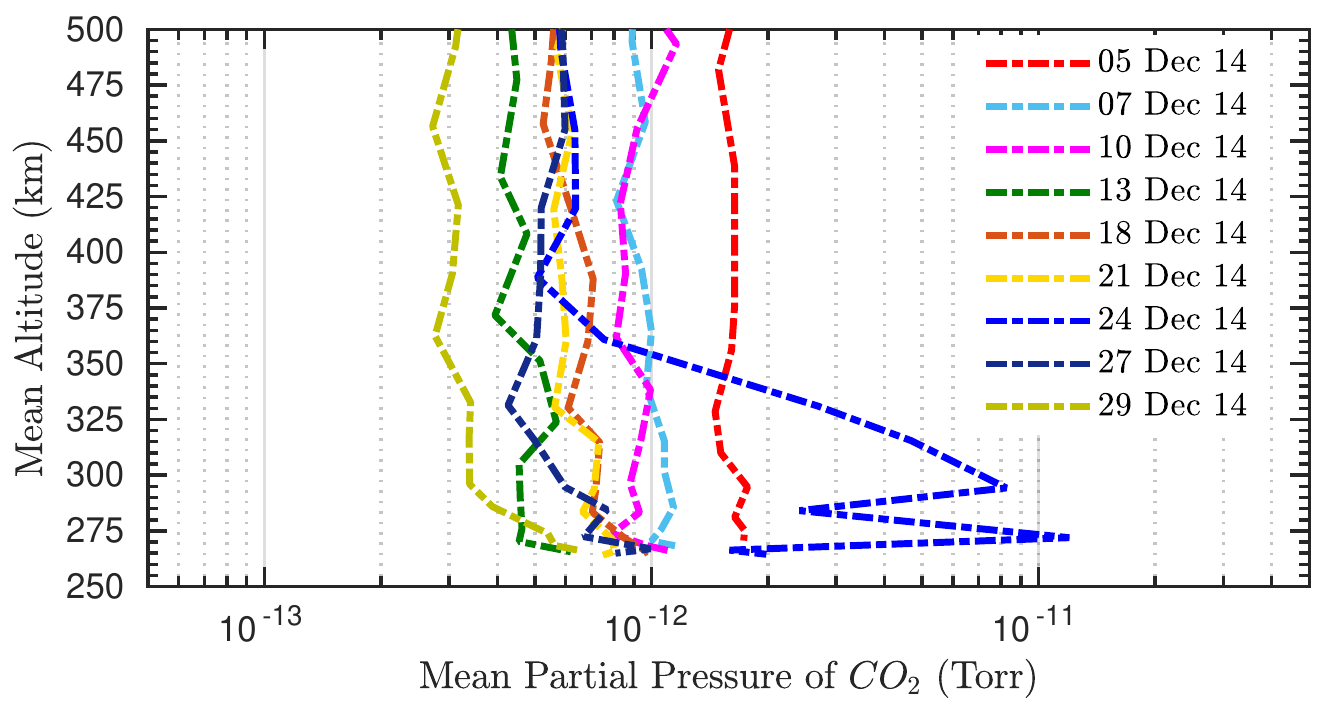}
	\end{minipage}
	\hfill
	\begin{minipage}[b]{0.4\paperwidth}
		\includegraphics[height=6cm,width=\textwidth]{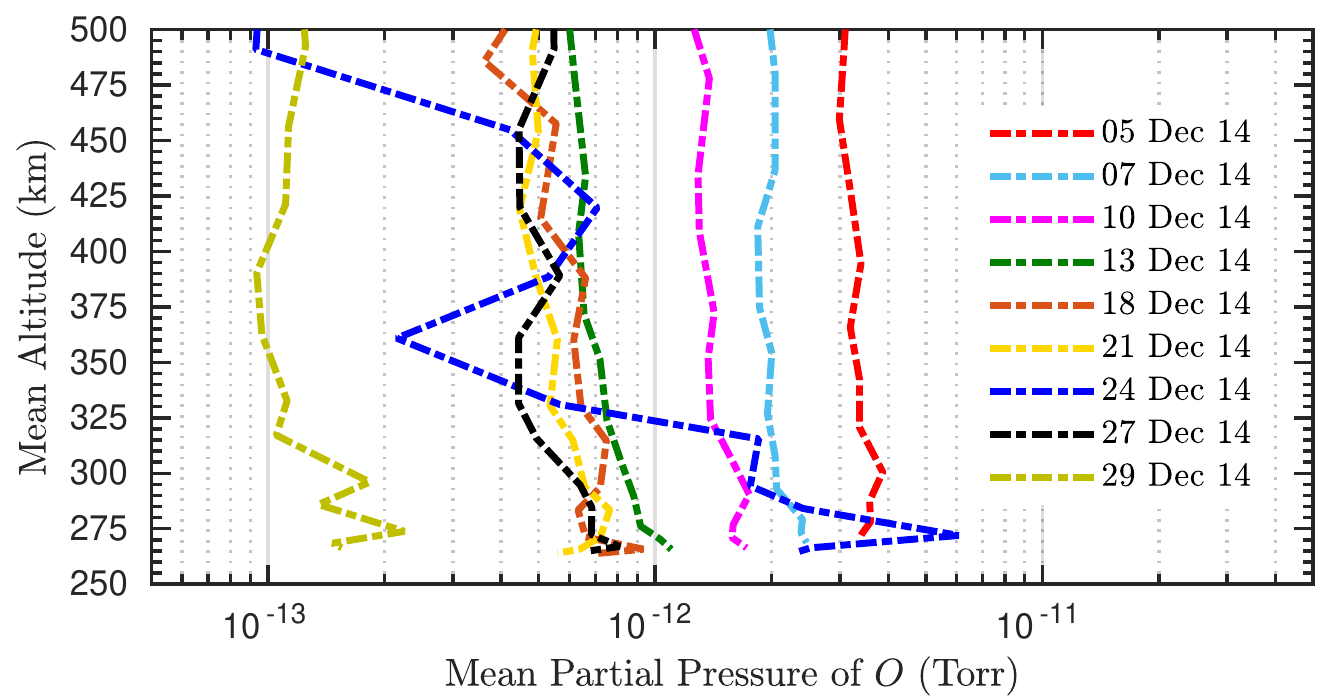}
	\end{minipage}
	\caption{MOM orbit-wise vertical profiles of partial pressure measured during December~2014\label{overflow}}
\end{figure}

\newpage

\begin{figure}[t]
	\centering
	\begin{minipage}[b]{0.4\paperwidth}
		\includegraphics[height=7cm,width=\textwidth]{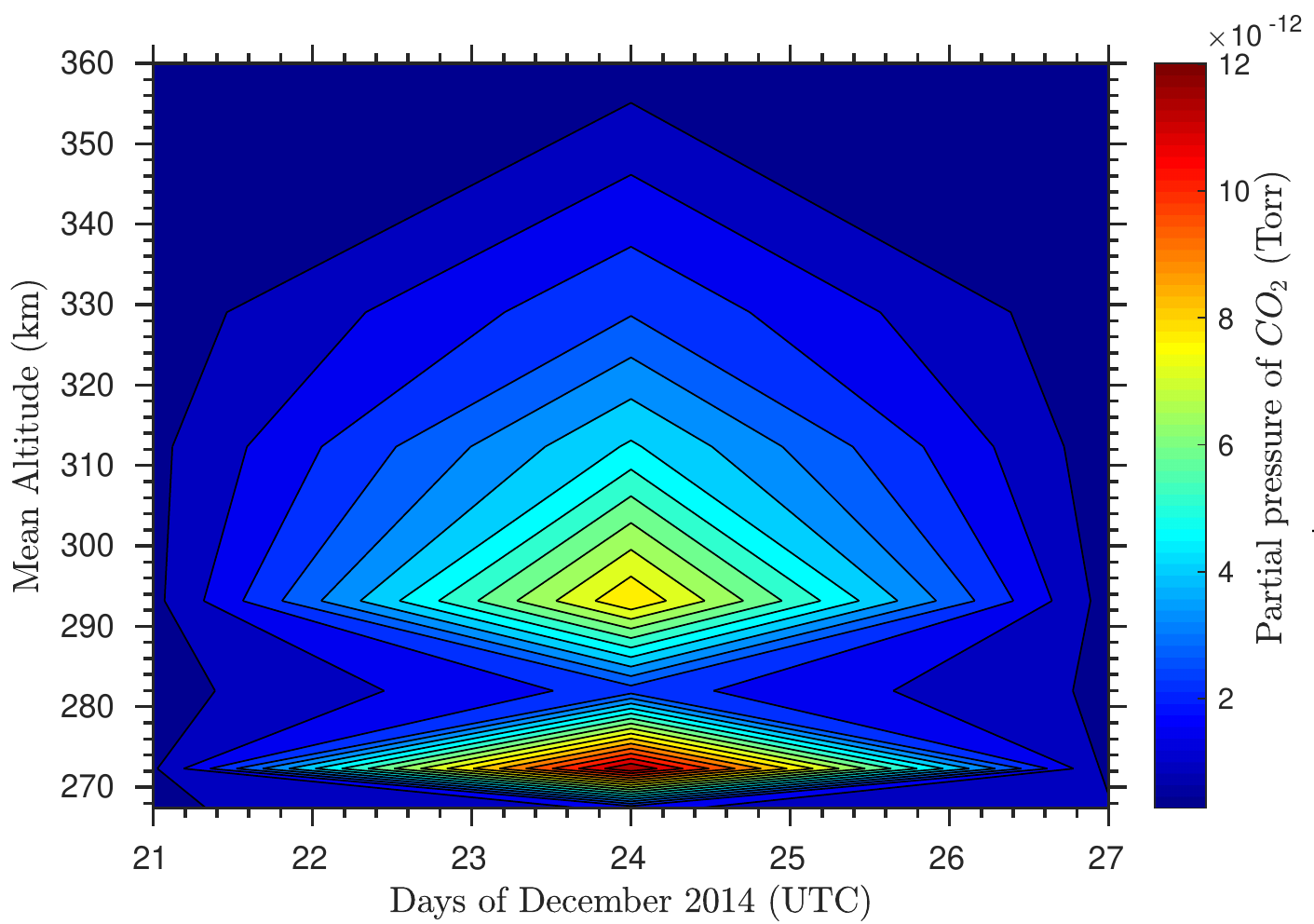}
	\end{minipage}
	\hfill
	\begin{minipage}[b]{0.4\paperwidth}
		\includegraphics[height=7cm,width=\textwidth]{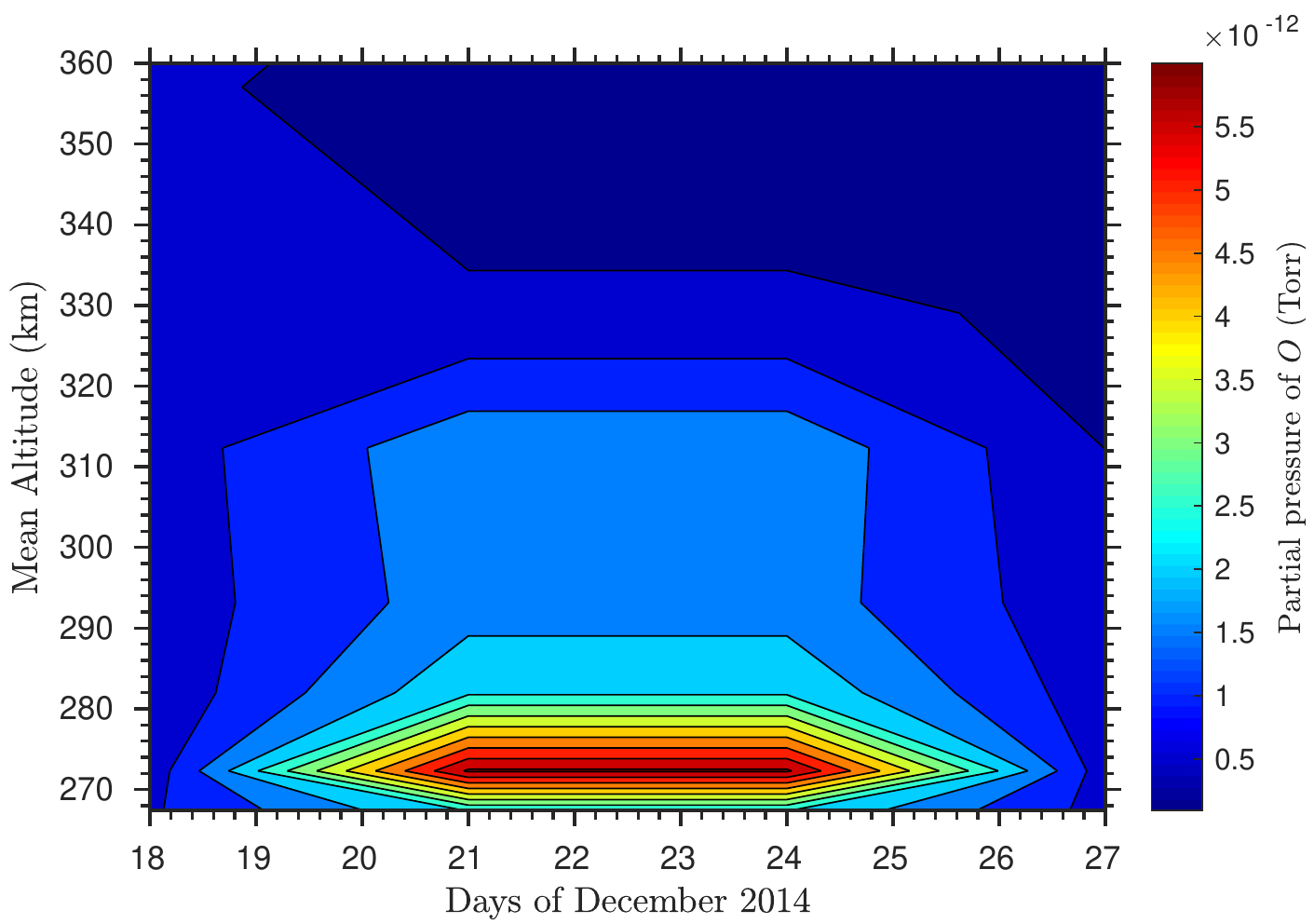}
	\end{minipage}
	\caption{Partial pressure contours of $CO_2$ during 21-27~December~2014 (left) and partial pressure contours of $O$ during 18-27~December~2014 from MENCA observations (right).}
\end{figure}

As can be inferred from the results presented in Figure 5 and 6, these anomalous effects are to be examined in terms of any eruptive events of solar high energy electromagnetic and charged particle radiations during December~2014 as both these radiations would arrive at Mars and interact with its atmosphere and surface. Since Mars does not have a magnetic field like that of Earth, in addition to the UV/X-rays, the charged particles or solar plasma components like energetic protons would directly interact with the neutral composition and initiate photodissociation and photoionisation. The absorption of particle energy would also result in modified densities of atmospheric species in certain height ranges due to increase in temperatures. To check this phenomenon, the solar activity as revealed by the smoothed daily mean SSN mentioned earlier, and the solar proton densities measured by Advanced Composition Explorer (ACE) spacecraft (srl.caltech.edu, 2018) are plotted for December~2014 and shown in Figure 11. 

\begin{figure}[h]
	\begin{center}
		\includegraphics[height=9cm,width=0.8\paperwidth]{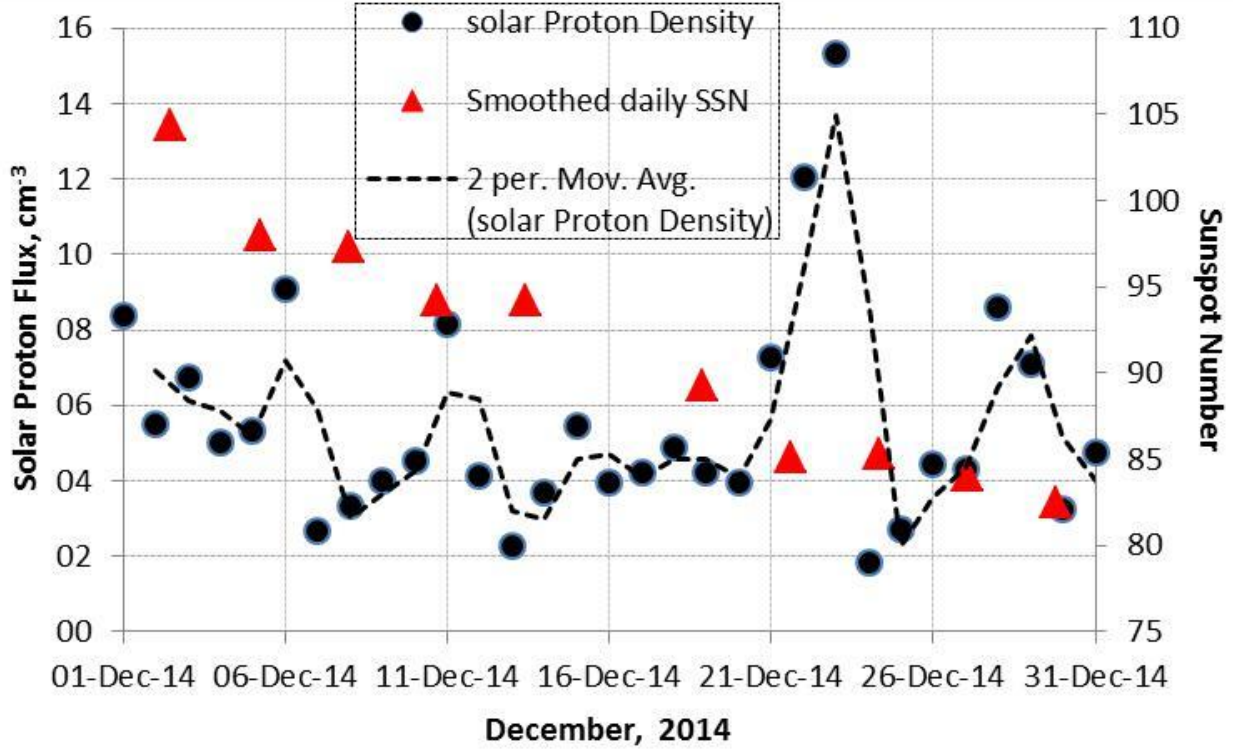}
		\caption{Variation of smoothed daily mean sunspot numbers and solar proton flux along with moving averages during December~2014}
	\end{center}
\end{figure}

The UV/X-rays fluxes for the period did not show any occurrence of solar flare during this period and hence are not shown in the figure. While it is known that CMEs, in large number of cases, are preceded by solar flares, it is also shown that this may not always be the case and occurrence of moderately strong CMEs during a very low solar activity year of 2009 has occurred without any associated solar flare (Nagaraja et al., 2018). Hence the variation of solar proton fluxes along with their moving averages in Figure 11 indicates that there is a large increase of energetic proton fluxes during the period 21-26~December~2014 with peak fluxes around 23~December~2014. The exact date of the peak of the corresponding $CO_2$ and $O$ pressures is not accurately discernable due to the fact that the observation periodicity is restricted to about 3-days orbital period of MOM. However it is clear the peaks of enhanced pressures of $CO_2$ has occurred with a delay of about 1 to 1.5 days. The $O$ pressures on the other hand show a broad maximum. The density of $O$ depends, apart ftom temperature on the photolysis and photo-ionisation effects and hence its enhancement and persistence of its peak values is primarily a cumulative effect compared to that of $CO_2$.

\section*{Summary and Conclusion}
\begin{itemize}
	\item MENCA data available from ISSDC in its nearly raw form has been processed to create a new data set with orbit-wise data assimilation particularly between 250 and 500 km for the period from September~2014 to September~2015. The ancillary information on the altitude, latitude, longitude and solar zenith angles obtained using the SPICE kernels have been tagged to each epoch of time and measurement. Also the partial pressures of the exospheric constituents of Mars have been converted to actual values using the calibration information and normalisation procedure.
	
	\item As a sample of altitude-pressure profiles of exospheric constituents $CO_2$, $O$, ($N_2$+$CO$), $O_2$,$NO$, $He$ \& $Ar$ obtained from MOM orbit $\#$55 near the periareion (covering $\approx$ 262-500 km in the exosphere of Mars) are generated and presented using this data set. The variation of profiles generally follows the exponential decrease with altitude with atomic Oxygen ($O$) concentrations being larger than that of $CO_2$ during daylight hours. 
	
	\item The total pressure is computed by summing up the partial pressures of all the major constituents i.e., $CO_2$, $O$, $N_2$, $CO$, $NO$, $O_2$, $Ar$, $N$ and $He$ to ensure removing those constituents affected by the degassing problem. The variation of total pressure at the periareion altitudes during October~2014 to May~2015 is well correlated with the smoothed daily sunspot numbers.
	
	\item A time series plot of partial pressure-altitude profiles of $CO_2$ and $O$ during December~2014 shows gradual decrease in partial pressure values (at all altitudes) from beginning to end of December which is again consistent with the decrease in the smoothed daily sunspot numbers.
	
	\item Superimposed on the December~2014 profiles of partial pressures of $CO_2$ and $O$ , there are anomalously high pressure profiles of both $CO_2$ and $O$ peaking on 24~December~2014. The contour plots show more details of the temporal and altitude build up, maximum and recovery phases of this anomaly. It is found that this anomaly is due to a moderately strong CME event of sun which peaked around 23~December~2014. 
\end{itemize}

\section*{Acknowledgement}
The authors are grateful to Indian Space Research Organisation (ISRO) for providing necessary funds to carry out this work under a research project vide reference ISRO:SPL:01.01.33/16. We acknowledge the use of data from the Mars Orbiter Mission (MOM), first inter-planetary mission of the Indian Space Research Organisation (ISRO), archived at the Indian Space Science Data Centre (ISSDC) and, NASA's Navigation and Ancillary Information Facility (NAIF) for the necessary SPICE kernels of Mars. Thanks are also due to Dr. Anil Bhardwaj, MENCA Principal Investigator and the payload team members notably, Dr. Smitha V. Thampi, Dr. T. P. Das, Dr. M. B. Dhanya, Space Physics Laboratory (SPL), Vikram Sarabhai Space Center for their valuable inputs and discussions. 

\section*{References}
\begin{description}
	
\item Acton, C.H., Ancillary Data Services of NASA's Navigation and Ancillary Information Facility, Planetary and Space Science, 44,1, 65-70, 1996 

\item Acton, C.H., Bachman, N., Semenov, B., Wright, E., A look toward the future in the handling of space science mission geometry, Planetary and Space Science, 2017 doi:10.1016/j.pss.2017.02.013

\item Acuna, M. H., Connerney, J. E. P., Ness, N. F., Lin, R. P., Mitchell, D., Carlson, C. W., McFadden, J., Anderson, K. A., Reme, H., Mazelle, C., Vignes, D., Wasilewski P., Cloutier, P., Global distribution of crustal magnetization discovered by the Mars Global Surveyor MAG/ER experiment, Science, 284(5415), 790-793, 1999 doi:10.1126/science.284.5415.790.

\item Barabash, S., Fedorov, A., Lundin, R., Sauvaud, J. A., Martian Atmospheric Erosion Rates, Science, 315(5811), 501-503, 2007 doi:10.1126/science.1134358

\item Barker, E. S., Detection of Molecular Oxygen in the Martian Atmosphere, Nature, 238, 447-448, 1972

\item Barth, C. A., and M. L. Dick, Ozone and polar hood on Mars, Icarus, 22, 205-201, 1974

\item Bhardwaj, A., Aliyas, A. V., Mohankumar, S. V., Das, T. P., Pradeepkumar, P., Sreelatha, P., Sundar, B., Amarnath, N., Dinakar, P. V., Dhanya, M. B., Neha, N., Supriya, G., Satheesh R. T., Padmanabhan, G. P., Vipin, K. Y., MENCA experiment aboard India’s Mars Orbiter Mission, Current Science, 109, 6, 1106-1113, 2015 doi:10.18520/v109/i6/1106-1113

\item Bhardwaj, A., Smitha, V. T., Das, T. P., Dhanya, M. B., Neha, N., Dinakar, P. V., Pradeepkumar, P., Sreelatha, P., Abhishek, J. K., Satheesh R. T., Vipin, K. Y., Sundar, B., Amarnath, N., Padmanabhan, G. P., Aliyas, A. V., Observation of Suprathermal Argon in the exosphere of Mars, Geophysical Research Letters, 44, 1-8, 2017 doi:10.1002/2016GL072001

\item Bhardwaj, A., Smitha, V. T., Das, T. P., Dhanya, M. B., Neha, N., Dinakar, P. V., Pradeepkumar, P., Sreelatha, P., Supriya, G., Mohankumar, S. V., Satheesh R. T., Vipin, K. Y., Sundar, B., Amarnath, N., Padmanabhan, G. P., Aliyas, A. V., On the evening time exosphere of Mars: Result from MENCA aboard Mars Orbiter Mission, Geophysical Research Letters, 43, 1862–1867, 2016 doi:10.1002/2016GL067707

\item Bougher, S. W., Cravens, T. E., Grebowsky, J., Luhmann, J., The aeronomy of Mars: Characterization by MAVEN of the upper atmosphere reservoir that regulates volatile escape, Space Sci Rev, doi:10.1007/s11214-014-0053-7, 2014

\item Bougher, S. W., Coupled MGCM-MTGCM Mars Thermosphere Simulations and Resulting Data Products in Support of the MAVEN Mission, JPL/CDP report, 1-9, 6 August 2012.

\item Carleton, N. P., and Traub, W. A., Detection of molecular oxygen on Mars, Science, 177, 988-992, 1972

\item Martínez, G. M., Newman, C. N., De Vicente-Retortillo, A., Fischer, E., Renno, N. O., Richardson, M. I., Fairen, A. G., Genzer, M., Guzewich, S. D., Haberle, R. M., Harri, A. M., Kemppinen, O., Lemmon, M. T., Smith, M. D., de la Torre-Juarez, M., Vasavada, A. R., The Modern Near-Surface Martian Climate: A Review of In-situ Meteorological Data from Viking to Curiosity, Space Sci Rev, doi:10.1007/s11214-017-0360-x, 295-338, 2017

\item Haberle, R. M., Early Mars climate models, J. Geophys. Res., 103(E12), 28467-28479, 1998 doi: 10.1029/98JE01396

\item Hansen, C. J., Thomas, N., Portyankina, G., McEwen, A., Becker, T., Byrne, S., Herkenhoff, K., Kieffer, H., Mellon, M., HiRISE Observations of Gas Sublimation-Driven Activity in Mars’ Southern Polar Regions: I. Erosion of the Surface, Icarus, 205, 283-295, 2010

\item Hanson, W. B., Sanatani, S., Zuccaro, D., The Martian ionosphere as observed by the Viking retarding potential analyzers, J. Geophys. Res, 82, 4351–4363, 1977 

\item Hanson, W. B., and Mantas, G. P., Viking electron temperature measurements: Evidence for a magnetic field in the Martian ionosphere, J. Geophys. Res., 93(A7), 7538–7544, 1988, doi: 10.1029/JA093iA07p07538.

\item Nagaraja, K., Praveen Kumar, B., Chakravarty, S. C., X-ray flares and coronal mass ejections (CMEs) during very quiet solar activity conditions of 2009, Ind. J. Pure Appl. Phys., 50, 621-623, 2018

\item Kaplan, L. D., Connes, J., Cannes, P., Carbon Monoxide in the Mars Atmosphere, Astro Q Phys J, 157, LI87-L192, 1969

\item Lane, A. L., Barth, C. A., Hord, C. W., Stewart, A. I., Mariner 9 Ultraviolet Spectrometer Experiment: Observations of Ozone on Mars, Icarus, 10, 102-l08, 1973

\item Magalhaes, J. A., Schofield, J. T., Seiff, A., Results of the Mars Pathfinder Atmospheric Structure Investigation, J. Geophys. Res., 104, 8943-8956, 1999
	
\item Mahaffy, P. R., Webster, C. R., Stern, J. C., Brunner, A. E., Atreya, S. K., Conrad, P. G., Domagal-Goldman, S., Eigenbrode, J. L., Flesch, G. J., Christensen, L. E., Franz, H. B., Freissinet, C., Glavin, D. P., Grotzinger, J. P., Jones, J. H., Leshin, L. A., Malespin, C., McAdam, A. C., Ming, D. W., Navarro-Gonzalez, R., Niles, P. B., Owen, T., Pavlov, A. A., Steele, A., Trainer, M. G., Williford, K. H., Wray, J. J., The imprint of atmospheric evolution in the D/H of Hesperian clay minerals on Mars, Science, 347 (6220), 412-414, 2015 doi:10.1126/science.1260291

\item Mangold, N., Quantin, C., Ansan, V., Delacourt, C., Allemand, P., Evidence for Precipitation on Mars from Dendritic Valleys in the Valles Marineris Area, Science, 305, 78-81, 2004

\item Moroz, V. I., Chemical composition of the atmosphere of Mars, Adv. Space Res., 22, 449-457, 1998

\item Nier, A. O., and McElroy, M. B., Composition and Structure of Mars’ Upper Atmosphere- Results from the Neutral Mass Spectrometers on Viking 1 and 2, J. Geophys. Res., 82, 4341-4349, 1977

\item Niles, P. B., Boynton, W., Hoffman, J. H., Ming, D. W., Hamara, D., Science, 329, 1334-1337, 2010

\item Olsen K., Montmessin, F., Fedorova A., Alexander T., Korablev, O., Trace gas retrievals for the ExoMars Trace Gas Orbiter Atmospheric Chemistry Suite mid-infrared solar occultation spectrometer, European Planetary Science Congress 2017, Riga, Latvia

\item Owen, T. S., Biemann, K., Rusbneck, D. R., Biller, L. E., Homarth, D. W., Lafleur, A. L., The Composition of the Atmosphere at the Surface of Mars, J. Geophys. Res. 82, 4635-4639, 1977

\item Mahaffy, P. R., Webster, C. R., Atreya, S. K., Franz, H., Wong, M., Conrad, P. G., Harpold, D., Jones, J. J., Leshin, L. A., Manning, H., Owen, T., Pepin, R. O., Squyres, S., Trainer, M., Abundance and Isotopic Composition of Gases in the Martian Atmosphere from the Curiosity Rover, Science, 341 (6143), 263-266, 2013 doi:10.1126/science.1237966

\item Withers, P., Lorenz R. D., Neumann G. A., Comparison of Viking Lander Descent Data and MOLA Topography Reveals Kilometre-Scale Offset in Mars Atmosphere Profiles, Icarus 159, 259-261, 2002 doi:10.1006/icar.2002.6914

\item Sagdeev, R. Z., and Zakharov, A. V., Brief history of the Phobos mission, Nature, 341(6243), 581-585, 1989, doi:10.1038/341581a0

\item Smith, M. D., THEMIS Observations of Mars Aerosol Optical Depth from 2002-2008, Icarus, 202, 444-452, 2009

\item Squyres, S. W., Arvidson, R. E., Bollen, D., Bell, J., Brückner, J., Cabrol, N. A., Calvin, W. M., Carr, M. H., Christensen, P., Clark, B. C., Crumpler, L., Des Marais, D. J., d'Uston, C., Economou, T., Farmer, J., Farrand, W. H., Folkner, W., Gellert, R., Glotch, T. D., Golombek, M. P., Gorevan, S., Grant, J. A., Greeley, R., Grotzinger, J., Herkenhoff, K. E., Hviid, S., Johnson, J. R., Klingelhöfer, G., Knoll, A. H., Landis, G., Lemmon, M. T., Li, R., Madsen, M. B., Malin, M. C., McLennan, S. M., McSween, H. Y., Ming, D. W., Moersch, J., Morris, R. V., Parker, T., Rice, Jr W., Richter, L., Rieder, R., Schröder, C., Sims, M., Smith, M., Smith, P., Soderblom, L. A., Sullivan, R. J., Tosca, N. J., Wänke, H., Wdowiak, T., Wolff, Michael J., Yen, Albert S. J. Geophy. Res. Planets, 3 (12), E12S12, 2006.

\item Garrard, T. L., Davis, A. J., Hammond, J. S., Sears, S. R., The ACE Science Center, Space Sci. Rev., 86, 1-4, 1998

\item Stephen R. L., Matthew C., Peter L. R., A climate database for Mars, J. Geophys. Res., 104, E10, 1-24, 1999

\item Young, L. D. G., and Young A. T., Interpretation of High-Resolution Spectra of Mars. IV. New Calculations of $CO$ abundance, Icarus, 30, 75-79, 1977
	
\end{description}

\end{document}